\documentstyle[amssymb,preprint,aps]{revtex}
\begin{document}
\title{Ordered droplet structures at the liquid crystal surface and
elastic-capillary colloidal interactions}
\author{I.I. Smalyukh$^{1}$, S. Chernyshuk$^{2}$, B.I. Lev$^{2,3}$, A.B. Nych%
$^{2}$, U. Ognysta$^{2}$, V.G. Nazarenko$^{2}$, and O.D. Lavrentovich$^{1}$ (%
$\thanks{%
Author for correspondence (e-mail: odl@lci.kent.edu)}$)}
\address{$^{1}$Liquid Crystal Institute and Chemical Physics\\
Interdisciplinary Program, Kent State University, Kent, OH 44242, USA\\
$^{2}$Institute of Physics, Prospect Nauki 46, Kyiv-39, 03039, Ukraine\\
$^{3}$Japan Science and Technology Corporation, Yokoyama Nano-structured\\
Liquid Crystal Project, 5-9-9 Tokodai, Ibaraki 300-2635, Japan}
\date{\today}
\maketitle

\begin{abstract}
We demonstrate a variety of ordered patterns, including hexagonal structures
and chains, formed by colloidal particles (droplets) at the free surface of
a nematic liquid crystal (LC). The surface placement introduces a new type
of particle interaction as compared to particles entirely in the LC bulk.
Namely, director deformations caused by the particles lead to distortions of
the interface and thus to capillary attraction. The elastic-capillary
coupling is strong enough to remain relevant even at the micron scale when
its buoyancy-capillary counterpart becomes irrelevant.
\end{abstract}

\pacs{61.30.-v, 82.70.Kj, 89.75.Fb, 68.03.Cd}

2D organization of nanometer- and micrometer-sized colloidal particles at
fluid interfaces is a fascinating phenomenon of both fundamental and applied
interest. \ The nature of micron-scale interparticle forces, especially of
the attractive nature, remains a subject of an ongoing debate \cite%
{1,2,3,4,5,7,9,10,11}. \ In most cases, the interparticle forces are
isotropic and cause hexagonal ordering. Anisotropic interactions can be
achieved when the particles are nonspherical \cite{4,5}. Apparently, {\it %
the surfaces of anisotropic fluids}, i.e., liquid crystals (LCs), can also
support anisotropic interactions and thus a richer variety of ordered
patterns as compared to isotropic fluids, but the current knowledge in the
field is rather limited. It is known that small particles might form chains
decorating the surface director field \cite{14,15} and that the surface air
bubbles can be accompanied by point defects \cite{16}. \ The behavior of
particles entirely in the LC\ bulk is studied much better \cite%
{R1,R4,R6,R7,R8,R9,R10,R11,R13}. \ \ Because of the anisotropy of molecular
interactions at the particle surface (the phenomenon of anchoring \cite{R17}%
), the embedded particle causes director distortions. \ The most frequently
met distortions are of dipole symmetry; they lead to chaining of droplets %
\cite{R1}. \ Recently, an unexpected and so far unexplained hexagonal
structures have been discovered for an array of glycerol droplets assumed to
be in the nematic bulk \cite{R9,R10,R11} and for droplets in smectic
membranes \cite{R13}.

In this work, we use confocal microscopy to demonstrate that placement of
colloidal particles (glycerol droplets of radius $R=1-10$ $\mu m$) {\it at
the LC surface} leads to attractive interactions and ordered patterns of
hexagonal and chain type that depend on the thickness of the LC film. \ The
attractive interactions can be explained by the elastic-capillary coupling,
as the particle-induced\ director deformations distort the LC surface. \ 

The glycerol droplets are obtained as in Ref.\cite{R9}. \ A Petri dish
containing a layer of glycerol and the nematic LC pentylcyanobiphenyl (5CB,
EM Industries) on top of it, is kept at $50^{0}C$ for 10 min, to facilitate
diffusion of the glycerol molecules into the 5CB layer which is in the
isotropic phase above $\ \approx 35^{0}C$. \ When the sample is cooled down,
solubility decreases and one observes appearance and growth of glycerol
droplets. \ The technique produces droplets of a constant radius $R$ \
determined by the cooling rate and the number of thermal cyclings. The
thickness of the LC film is controlled in the range $h=(3-100)$ $\mu m$. The
5CB film is in the hybrid aligned state: the director ${\bf \hat{n}}$ is
parallel to the LC-glycerol interface at the bottom and perpendicular to the
air-LC interface at the top.

In order to determine the location of droplets, we use the fluorescence
confocal\ polarizing microscopy (FCPM) \cite{R20}. Two different dyes with
separated absorption and fluorescence bands, fluorescein and Nile red
(Aldrich), were added $(0.01$ wt $\%)$ to tag glycerol and LC, respectively %
\cite{R20}; they did not change the appearance of patterns. \ The FCPM
textures of vertical cross-sections, Fig.1a-c, unambiguously demonstrate
that the glycerol droplets are trapped at the LC-air interface. For a better
clarity, the images in Fig.1a,b are taken for a large drop; smaller drops
are also located at the interface, Fig.1c.

A 4-6 hour relaxation at room temperature results in 2D ordered structures
of droplets. \ The order is hexagonal when the LC layer is thick, $h>>R$, $%
h=\left( 20-100\right) \mu m$, Fig.2a,b. \ The droplet center-to-center
separation $r_{eq}$ increases with the droplets diameter $2R$, Fig.2d; each
data point is an average of the data collected by video-recording the
patterns every 6 s for 10 min. The angles between the directions to the
nearest neighbours changes in the range $60^{0}\pm 8^{0}$ for
well-equilibrated structures. \ At this stage, it is hard to quantify the
lattice symmetry more precisely as it might be influenced by slowly changing
director pattern in the bulk. In thin films, $h/R\approx 2\div 5$, $3\mu
m\leq h\leq 10\mu m$, Fig.3a, the pattern changes completely, as the
droplets form chains oriented along the average horizontal projection of $%
{\bf \hat{n}}$. The length of chains (number of droplets in a chain)
increases when $h$ decreases. Importantly, both hexagonal and chain patterns
disorganize when 5CB is heated into the isotropic phase.

The very existence of 2D hexagonal pattern in thick films does not
discriminate between attractive and repulsive interactions among the
particles, as in the confined systems such as a Petri dish the ordering can
be induced by purely repulsive forces. The presence of repulsive force is
manifested clearly, as the droplets are well separated, $r_{eq}\approx
\left( 3-4\right) R$, Fig.2d. \ The following three experiments show that
the interaction has also an attractive component. (A) We start with the
Petri dish in which the 2D lattice occupies the whole area and then use a
filter paper to wipe about 1/2-1/3 of the droplets. \ The remaining droplets
restore the 2D lattice within the smaller area, with the lattice parameter
close to the original one.\ (B) We change the direction of the meniscus tilt
near the Petri dish walls by coating the walls with a hydrophobic material
(stearin); the types of the 2D ordering remain the same. \ (C) After the 2D
lattice is formed in a regular manner, we add (through a microsyringe) a
small amount (0.01\% by weight) of a surfactant (cetylpiridinium chloride)
to the nematic layer. \ The interparticle distances decrease; the shrinked
clusters are separated by droplet-free areas, Fig.2c. \ 

What is the mechanism of the droplet attraction? The gravity-mediated
capillary effect and the van der Waals forces appear to be insignificant, as
the corresponding pair potentials \cite{R21,R17}

\begin{equation}
U_{g}\left( r\right) \approx \frac{2\pi R^{6}}{\sigma _{LCA}}g^{2}\rho
_{LC}^{2}\ln (\frac{r}{\lambda })\text{; \ \ \ \ \ \ \ \ \ \ \ }%
U_{vdW}\left( r\right) \approx -\frac{AR}{12\left( r-2R\right) };  \label{1}
\end{equation}%
are of the order of thermal energy $k_{B}T\approx 4\times 10^{-21}$\ $J$ or
less, when $r$ is a few microns. Here $\sigma _{LCA}\approx 3.8\ast
10^{-2}J/m^{2}$ \cite{R21A} is the LC-air surface tension coefficient, $\rho
_{LC}$ $\approx 10^{3}$ $kg/m^{3}$ is the LC density, $g$ is acceleration
due to the gravity, $\lambda =\sqrt{\sigma _{LCA}/\left( g\rho _{LC}-\frac{%
\partial \Pi }{\partial h}\right) }$ is the capillary length; $\Pi $ $%
\approx K/h^{2}$ is the elastic disjoining pressure for the hybrid aligned
film \cite{R17}; $K\sim 10^{-11}N$ is the average nematic elastic constant;
in all cases of interest to us, $\lambda >>r_{eq}$, as $\lambda \approx 2mm$
when $h\rightarrow \infty $ and $\lambda \approx 0.5mm$ when $h=5\mu m$;
finally, $A$ is the Hamaker constant of the order of $10^{-19}J$ or smaller %
\cite{R17}.

Attraction of droplets at the nematic surface (and absence of attraction and
ordering when 5CB is in the isotropic state) can be explained by the
coupling of orientational elasticity and capillarity. \ The elastic
distortions caused by the submerged part of the particle give rise to a
vertically resolved force $f_{el}$ that shifts the LC-air interface upward
or downward,\ depending on the particle properties, to reduce the elastic
energy of the LC host; $f_{el}$ is balanced by surface tension. Non-flat
interface causes long-range attraction of particles. \ As the first
approximation, we divide the total interaction potential $U\left( r\right) $
into two parts, $U_{elc}\left( r\right) $ describing the elastic-capillary
coupling and $U_{elb}\left( r\right) $ describing the \textquotedblright
pure bulk\textquotedblright\ elastic interactions. \ 

To estimate the orders of magnitude, we use the idea that $f_{el}$ is
related to the non-vanishing surface anchoring of ${\bf \hat{n}}(r)$ at the
LC-particle interface: $f_{el}=\frac{\partial }{\partial z}E_{el}(z)\sim 
\frac{\partial }{\partial z}\oint_{\Omega (z)}W({\bf \hat{n}\cdot \hat{\nu}}%
)^{2}ds,$ where $E_{el}(z)$ is the total elastic energy of distortions
caused by the particle, $\Omega (z)$ is part of the droplet surface immersed
in the LC, $W$ is the (polar) anchoring coefficient of the LC-glycerol
interface, ${\bf \hat{\nu}}$ is the normal to the interface, and $z$ is the
vertical coordinate of some fixed point, say, the bottom of the droplet. For 
$f_{el}\approx 4\pi WR$, with $W=10^{-5}J/m^{2}$ \cite{R17}, one finds $%
f_{el}\sim 10^{-10}N\sim 10K$ when $R=1\mu m$. The force $f_{el}$ replaces
the buoyancy force in the standard treatment \cite{3,R21}; it is balanced by
the surface tension, hence, $f_{el}=2\pi \sigma _{LCA}r_{i}\sin \psi $,
where $\psi $ is the meniscus slope at the triple contact line of radius $%
r_{i}$. \ The vertical displacement $\xi \left( r\right) \approx -r_{i}\sin
\psi \ln \frac{r}{\lambda }$ changes the interfacial area and leads to the
attractive potential written for two identical particles as $%
U_{elc}=-f_{el}\xi $, or

\begin{equation}
U_{elc}\left( r\right) =\frac{f_{el}^{2}}{2\pi \sigma _{LCA}}\ln (\frac{r}{%
\lambda }),  \label{3}
\end{equation}%
similar to the buoyancy case \cite{3,R21}. For $f_{el}\sim 10K,$ one
estimates $U_{elc}\sim 3\times 10^{-19}J\sim 80$ $k_{B}T$.\ The
elastic-capillary coupling is strong enough to keep droplets attracted even
at micron scales where the gravity mechanism vanishes: the dependence $%
U_{elc}$ $\left( R\right) $ is weaker than $U_{g}\left( R\right) \varpropto
R^{6}$ in Eq.(\ref{1}), as $E_{el}\left( R\right) \sim WR^{2}$ for $R<<K/W$,
and $E_{el}\left( R\right) \sim KR$ for $R>>K/W$ \cite{R17}.

We turn now to the \textquotedblright pure bulk\textquotedblright\ elastic\
interactions. \ In the hybrid aligned film, ${\bf \hat{n}}$ is determined by
the balance of anchoring and elasticity. In the thick films ($h>>K/W$, $h>>R$%
), ${\bf \hat{n}}$ at the LC-air interface is vertical. Tangential anchoring
of ${\bf \hat{n}}$ at the droplet surface leads to a point defect-boojum %
\cite{R17} close to the south pole, Fig.1e and 2e. The distortions ${\bf 
\hat{n}}\left( {\bf r}\right) $ around each droplet can be approximated by
an elastic dipole normal to the LC-air interface \cite{R1},\ ${\bf p}\approx
\left( 0,0,\alpha _{z}R^{2}\right) $, where $\alpha _{z}$ is a constant. The
bulk elastic dipole-dipole interaction is thus isotropic in the $xy$ plane
and repulsive,

\begin{equation}
U_{elb,thick}\left( r\right) =K\frac{\alpha _{z}^{2}R^{4}}{r^{3}}.
\label{ref4}
\end{equation}%
The total pair potential, $U=U_{elc}+U_{elb,thick}$, has a clear minimum at

\begin{equation}
r_{eq}^{pair}=\left( \frac{6\pi \alpha _{z}^{2}\sigma _{LCA}R^{4}K}{%
f_{el}^{2}}\right) ^{1/3};  \label{7}
\end{equation}%
e.g., $r_{eq}^{pair}=13\mu m$ for $f_{el}=10^{-10}N$ , $\alpha _{z}=0.2$ and 
$R=3\mu m$. For $f_{el}\varpropto WR$, Eq.(\ref{7}) predicts $%
r_{eq}^{pair}=\gamma \left( \sigma _{LCA}R^{2}K/W^{2}\right)
^{1/3}\varpropto R^{2/3}$, where $\gamma $ is a dimensionless constant. The
trend is in agreement with the experiment, Fig.2d, despite the fact that Eq.(%
\ref{7}) is derived from the pair potential. \ Generally, $f_{el}$ might be
a non-linear function of $R$ and also might depend on $r$. \ A surfactant
can be used to control the patterns, as it changes $\sigma _{LCA}$,
anchoring at interfaces, and thus $f_{el}$. Sometimes, one observes the
patches of a hexagonal \textquotedblleft dense\textquotedblright\ pattern
with $r_{eq,d}\approx \left( 2.2-2.7\right) R$ that coexist with the regular
hexagonal patterns of a larger period $r_{eq}\approx \left( 3-4\right) R$
(but the same $R$). \ These dense patches are not understood yet, and might
be provoked by ionic and flexoelectric charge, by the $r$-dependence of $%
f_{el}$, \ surfactant contamination, etc.

In the thin films, ${\bf \hat{n}}$ becomes progressively horizontal at the
LC-air interface as the anchoring coefficient $W_{air}$ there is smaller
than $W$ \cite{R23}. When $h$ decreases to $2R$ and $K/W_{air}$, the boojum
shifts to the side, Fig.3b, giving rise to the in-plane elastic dipole
component, $p_{x}=\alpha _{x}R^{2}$, and an anisotropic interaction
contribution

\begin{equation}
U_{elb,thin}\left( r,\theta \right) =K\frac{\alpha _{x}^{2}R^{4}}{r^{3}}%
\left( 1-3\cos ^{2}\theta \right) ,  \label{5}
\end{equation}%
that depends on the angle $\theta $ between the radius-vector connecting the
two droplets and the direction of $p_{x}$'s. The chain formation can thus be
explained similarly to the chaining of droplet-hedgehog pairs entirely in
the LC bulk \cite{R1}: $U_{elb,thin}$ is minimum when $\theta =0$. The
analogy is not complete, as our problem is of a lower symmetry and requires
one to consider $h$-, $W_{air}$-, and $W$-dependences of ${\bf \hat{n}(r)}$,
etc. For example, if the film is so thin that ${\bf \hat{n}}$ is parallel to
the LC-air interface \cite{R23}, then ${\bf \hat{n}(r)}$ around the droplets
might become of a quadrupolar type with two boojums along the horizontal
axis.

To conclude, the particles (glycerol droplets) at the LC-air interface are
capable of both repulsive and attractive interactions that lead to ordered
patterns of hexagonal and chain types. The director distortions in LC give
rise to a vertical elastic force balanced by surface tension; the deviation
of the LC-air interface from the horizontal plane results in attraction of
particles. The phenomenon is general, as we observe hexagonal patterns not
only for the liquid droplets but also for micron-size polymer spheres at the
nematic-air interface; these results will be published elsewhere. \ The
elastic-capillary mechanism can also shed some light at the observations of
hexagonal patterns claimed to form in the bulk of the hybrid aligned films %
\cite{R9,R10,R11}. \ If the particles were indeed in the bulk, where ${\bf 
\hat{n}}$ is tilted, then the elastic interaction potential would be
anisotropic and thus generally inconsistent with the hexagonal order. The
elastic model \cite{R9} does not consider the angular dependence of the
potential for tilted ${\bf \hat{n}}$ and does not explain the hexagonal
ordering. A pure elastic model would not be satisfactory even for thick
films, in which ${\bf \hat{n}}$ around the droplets is vertical, as in this
case the interaction would be only repulsive, Eq.(\ref{ref4}). \ Note that
the experiments \cite{R9,R10,R11} relied on the standard (non-confocal)
microscopy with a poor $z$-axis resolution. The particles might have been
located at the interface rather than in the LC bulk, which would bring the
observations \cite{R9,R10,R11} into the category of elastic-capillary
phenomenon proposed in this work.

We acknowledge helpful discussion with M. Kleman, V. Pergamenshchik, S.V.
Shiyanovskii, P. Tomchuk, and H. Yokoyama. The work was supported by
National Science Foundation Grant DMR-0315523 and STCU grant \#2025. ODL
thanks the participants of the "Geometry and Materials Sciences" Program at
the Aspen Center for Physics for discussions.

\newpage

{\LARGE Figure captions.}

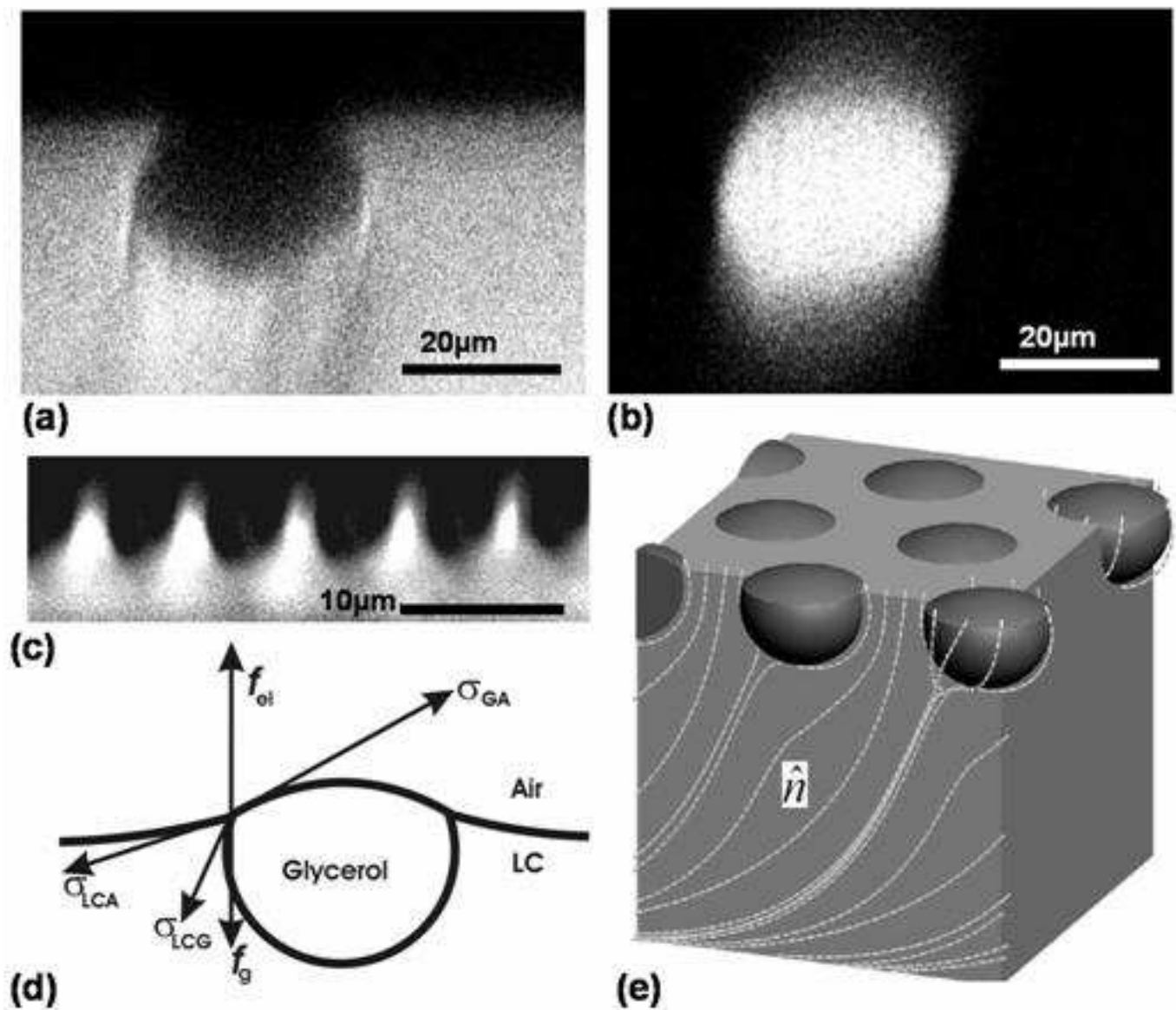
FIG.1. Two-channel FCPM\ textures of the vertical cross-section of the LC
film: (a,b) a single glycerol droplet at the LC-air interface; (c) a raw of
droplets that is a part of a hexagonal pattern at the LC-air interface; (d)
scheme of forces and interfacial tension vectors at the LC-air interface
with a strongly exaggerated meniscus slope; (e) 3D director field in the LC
layer. The LC layer (glycerol droplet) is manifested by the high (low)
intensity of fluorescent light in parts (a,c); the contrast is opposite in
(b).

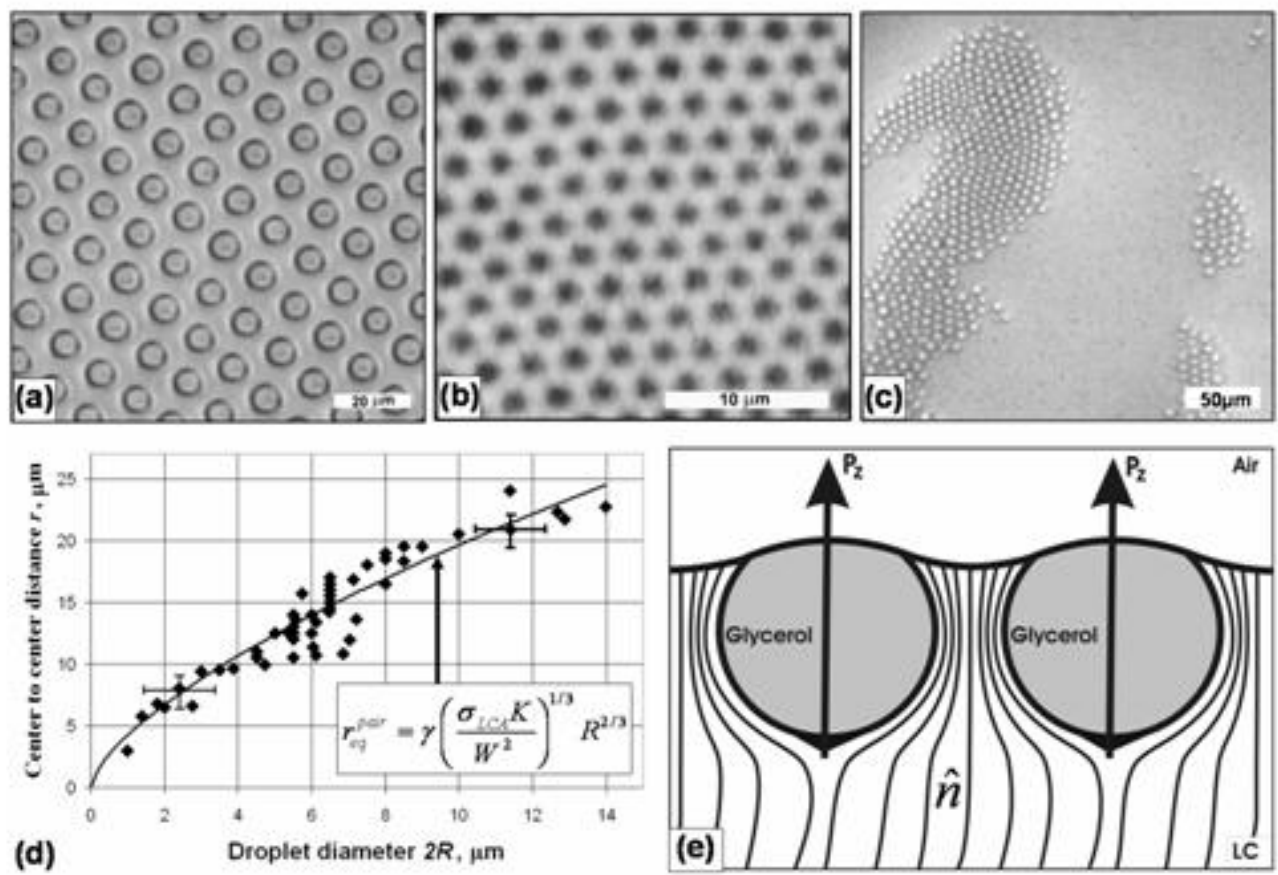
FIG.2. Optical microscopy pictures of hexagonal structures at the free
surface of the thick nematic layer, $h\approx 60\mu m$, formed by droplets
of different average diameter: (a) $2R\approx 7\mu m$, (b) $2R\approx 1\mu m$%
. (c); arrangement of droplets after the surfactant is added to the nematic
layer; (d)\ droplets separation $r_{eq}$ vs. $2R$; the data are compared to
the dependence $r_{eq}^{pair}=\gamma \left( \sigma _{LCA}R^{2}K/W^{2}\right)
^{1/3}$, where $\gamma =0.43$, $K=10^{-11}N$, $\sigma _{LCA}=3.8\times
10^{-2}J/m^{2}$, $W=10^{-5}J/m^{2}$, see the text; (e) schematic drawing of
the director field in the top part of the LC layer.

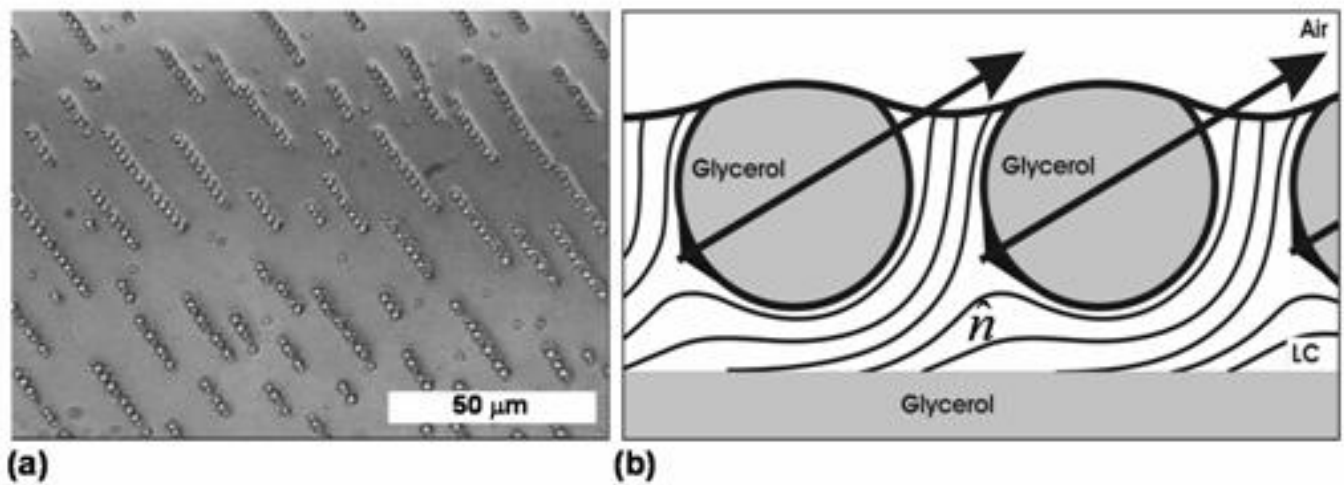
FIG.3. Optical microscopy textures of glycerol droplets\ forming chains at
the free surface of a thin LC layer, $h\approx \left( 7-10\right) \mu m$ (a)
and the corresponding director field scheme (b).

\end{document}